\documentclass{aastex}          
\usepackage{spr-astr-addons}    

\begin{document}
%
\title{A Bayesian Approach to Calibrating Period-Luminosity Relations of RR Lyrae Stars in the Mid-Infrared}

\shorttitle{Bayesian PL Relations}
\shortauthors{C. R. Klein et al.}

\author{Christopher R. Klein\altaffilmark{1}}
\and \author{Joseph W. Richards\altaffilmark{2}}
\and \author{Nathaniel R. Butler\altaffilmark{3}}
\and \author{Joshua S. Bloom\altaffilmark{1}}

\email{cklein@astro.berkeley.edu}

\altaffiltext{1}{Astronomy Department, University of California, Berkeley, CA 94720, USA}
\altaffiltext{2}{Statistics Department, University of California, Berkeley, CA 94720, USA}
\altaffiltext{3}{School Of Earth and Space Exploration, Arizona State University, Tempe, AZ 85287, USA}

\begin{abstract}
A Bayesian approach to calibrating period-luminosity (PL) relations has substantial benefits over generic least-squares fits. In particular, the Bayesian approach takes into account the full prior distribution of the model parameters, such as the {\it a priori} distances, and refits these parameters as part of the process of settling on the most highly-constrained final fit. Additionally, the Bayesian approach can naturally ingest data from multiple wavebands and simultaneously fit the parameters of PL relations for each waveband in a procedure that constrains the parameter posterior distributions so as to minimize the scatter of the final fits appropriately in all wavebands. Here we describe the generalized approach to Bayesian model fitting and then specialize to a detailed description of applying Bayesian linear model fitting to the mid-infrared PL relations of RR Lyrae variable stars. For this example application we quantify the improvement afforded by using a Bayesian model fit. We also compare distances previously predicted in our example application to recently published parallax distances measured with the Hubble Space Telescope and find their agreement to be a vindication of our methodology. Our intent with this article is to spread awareness of the benefits and applicability of this Bayesian approach and encourage future PL relation investigations to consider employing this powerful analysis method.
\end{abstract}


%

\section{Introduction}\label{s:intro}

The period--luminosity (PL) relations of pulsating variable stars --- typically variables of types RR Lyrae (RRL), Cepheid, and Mira --- are invaluable tools for constructing the rung of the distance ladder that connects the Milky Way to other nearby galaxies, extending to $\sim$ 5 Mpc. Recent applications of this distance measurement technique using Cepheids have successfully mated Cepheid distances to SNe Ia host galaxies and constrained the Hubble Constant, $H_0$, to 3.3\% \citep{2011ApJ...730..119R}. The authors have recently derived mid-infrared PL relations for RRL variables \citep{2011ApJ...738..185K}, and demonstrated their potential to serve as important distance indicators for the Large Magellanic Cloud. Additionally, continuing studies of Miras \citep{2008MNRAS.386..313W} confirm their potential to provide accurate distances even beyond the reach of Cepheids.

The accuracy and precision of any distance measurement made using the PL relation of a variable star, or any population of variable stars within a distant system, is dominated by the uncertainty of the locally calibrated PL relation. The general method is to fit a PL relation to the variables for which trigonometric parallax measurements are available \citep{1997MNRAS.286L...1F}. For more than the past decade only {\it Hipparcos} (original catalog published as \cite{1997ESASP1200.....P}, and improved reduction by \cite{2007ASSL..350.....V}) could provide these required local distance measurements to a significantly large sample of local stars with the accuracy necessary. More recently, the {\it Hubble Space Telescope} Fine Guidance Sensor has been used to provide higher accuracy parallax measurements for nine Cepheids \citep{2007AJ....133.1810B} and five RRL variables \citep{2011AJ....142..187B}. In the coming decade, with the launch of the ESA's astrometry mission Gaia, the sample size of potential PL relation calibrators and the accuracy of their parallax distances will be significantly augmented \citep{2009MmSAI..80...97C}.

PL relations are typically calibrated using straightforward, simple, frequentist statistical techniques (see, for example, \cite{2006MNRAS.372.1675S} fitting RRL variables, \cite{2006MNRAS.370.1979M} fitting type II Cepheids, \cite{1997MNRAS.286L...1F} fitting classical Cepheids, and \cite{2003MNRAS.343...67G} fitting Miras). At the basic level, a PL relation is an equation of the form $M=\alpha \log P + \beta$, where $M$ is the absolute magnitude (in a given waveband), $\alpha$ is the slope, $P$ is the period (in days), and $\beta$ is the zero point magnitude (which itself may be a function of metallicity). This simple linear equation can be reliably fit with the method of least squares, and the accuracy of the fit can be assessed with the standard deviation of the residuals, a metric commonly referred to as the scatter.

A significant limitation of the least squares regression method is that it does not make use of the full prior probability distribution of the parallax distances. Allowing the distances more flexibility to move within their prior distributions translates into posterior distances that are more consistent with the fitted PL relation and therefore more accurate (on average) than the prior distance mean. A second limitation is that this traditional method is not easily adapted to fitting PL relations of the same variables in different wavebands simultaneously. Intelligently combining data from multiple wavebands has the potential to produce better final PL relation fits.

A Bayesian approach for fitting the PL relation parameters overcomes these traditional limitations. \cite{2003ApJ...592..539B} discusses in substantial depth the application of a Bayesian approach to the Cepheid distance scale, using physical pulsational models and radial velocity data. In the present work we confine our examination to the application of Bayesian methods in calibrating, purely phenomenologically, the PL relations of pulsating variable stars. We use as our example the calibration performed by \citet{2011ApJ...738..185K}. In Section \ref{s:tech} we describe the generalized Bayesian modeling approach. In Section \ref{s:app} we work through the application of this Bayesian approach to mid-infrared PL relations of RRL variables, as first demonstrated by the authors in \cite{2011ApJ...738..185K}. We perform a traditional, least squares fit to the RRL PL relations and compare with the fits from our Bayesian approach in Section \ref{s:comp}. Finally, in Section \ref{s:conc} we draw conclusions and discuss future applications.

\section{Technical Explanation of Bayesian PL Relation Fitting}\label{s:tech}

An excellent and thorough description of Bayesian fitting of linear models is provided in \cite{gelman2003}. \cite{2003ApJ...592..539B} applies Bayesian analysis to the Cepheid distance scale, but does not use linear Bayesian model fitting for deriving the Cepheid PL relation. Here we review the foundation of the Bayesian approach.

If we assume that our data, denoted by $y$, follows some pattern or rule or model, as in common in nature, and denote the model parameter(s) by $\theta$, then we may write the probability of the model being true as $p(\theta)$ and the posterior probability that the model is true given our observed data $y$ as $p(\theta | y)$. The probability $p(\theta)$ is the prior distribution on the model, it is what we know before making observations. We can also define the likelihood as the probability $p(y|\theta)$ of observing the particular data $y$ conditioned on the model $\theta$. Thus, we have the unnormalized Bayes' theorem 
\begin{equation} \label{eqn:bayes_1}
p(\theta | y) \propto p(\theta) p(y|\theta),
\end{equation}
the probability of the model being true given the data is proportional to the prior probability of the model times the likelihood\footnote{The exact (normalized) form of Bayes' theorem is 
\begin{equation} \label{eqn:bayes_2}
p(\theta | y) = \frac{p(\theta) p(y|\theta)}{p(y)},
\end{equation}
but for our purposes we implement Equation \ref{eqn:bayes_1} by Monte Carlo computer simulation and thus the analytical normalization can be ignored.}.

To fit observed data $y$ to a model with parameters $\theta$ we simply evaluate Equation \ref{eqn:bayes_1} throughout a fine grid of values for $\theta$ to create the posterior distribution of the model. We can examine this posterior distribution (through analysis of repeated random draws from the distribution) to find the most likely fit parameters, as well as uncertainty in these parameters. Furthermore, this posterior distribution reveals the most likely true values for the data $y$, which is of course conditional on the prior distributions of the data. 

\section{Application to Mid-Infrared RRL Variables}\label{s:app}

In \cite{2011ApJ...738..185K} we apply Bayesian model fitting to a sample of 76 RRL light curves observed with the Wide-Field Infrared Survey Explorer (WISE) \citep{2010AJ....140.1868W}. Each of the RRL variables was well-observed in three WISE bands (W1 at 3.4, W2 at 4.6, and W3 at 12 $\mu$m) and their prior distance distributions are generated by applying the RRL $M_V$--[Fe/H] relation given in \cite{postHipp..book} to their {\it Hipparcos} mean flux $V$-band magnitudes, correcting for dust extinction.

Using the nomenclature of Section \ref{s:tech} we define our observed data for each RRL as $y=(m, P)$, the apparent magnitude and period. The unknown fit parameters are then $\theta = (\mu, M_0, \alpha, \sigma)$, the distance modulus, absolute magnitude zero point, PL relation slope, and scatter. We can also define $\beta = (\mu, M_0, \alpha)$ so that then $\theta = (\beta, \sigma)$. We put an informative normal prior on each $\mu$ (from the $V$-band distance estimates) and we put a flat prior on everything else.

We then write our statistical model of the PL relationship as
\begin{equation} \label{eqn:model}
m_{ij} = \mu_i + M_{0,j} + \alpha_j \log_{10}(P_i/P_0) + \epsilon_{ij},
\end{equation}
where $\mu_i$ is the distance modulus for $i$th RRL, $M_{0,j}$ is the absolute magnitude zero point for the $j$th WISE
band at $P=P_0$, where $P_0 = 0.50118$ day is the mean period of the sample, and $\alpha_j$ is the slope of the PL 
relationship in the $j$th band. We assume that any extinction is negligible in these bands. The error terms $\epsilon_{ij}$
are independent zero-mean Gaussian random deviates with variance $(\sigma \sigma_{m_{ij}})^2$, which describe the intrinsic scatter in the $m_{ij}$ about the model, where $\sigma$ is a free parameter which is an unknown scale factor on the known measurement errors, $\sigma_{m_{ij}}$\footnote{The average measurement error, $\sigma_m$, is 0.013, 0.013, and 0.045 mag in W1, W2, and W3, respectively.}. We fit the model (Equation \ref{eqn:model}) using a 
Bayesian procedure, outlined below and explicitly described in Section 4 of \cite{2011ApJ...738..185K}.

First, we assume a normal (Gaussian) prior distribution on
each of the distance moduli with mean $\mu_{0,i}$ and standard deviation $\sigma_{\mu_{0,i}}$, as described above.
For the other parameters in our model (Equation \ref{eqn:model}), we assume a
flat, noninformative prior distribution.  

Our likelihood is normal: 
\begin{equation} \label{likelihood}
p(m,P | \beta,\sigma) \propto p(m | \beta,\sigma) = N(\mathbf{X} \beta,\sigma^2 \textrm{diag}(\sigma_m^2)),
\end{equation}
where $N$ denotes the multivariate normal distribution.
Note that $p(m,P | \theta) \propto p(m | \theta)$ since $P$ is independent of $m$ and doesn't depend on any of the parameters $\theta$.

In the nomenclature of Section \ref{s:tech} we are solving for 
\begin{equation}\label{eqn:posterior}
p(\theta | y) = p(\beta, \sigma | m,P) = p(\beta | m,P,\sigma)  p(\sigma | m,P).
\end{equation}
The joint posterior distribution, $p(\beta,\sigma | m,P)$, can be sampled by
first drawing from $p(\sigma | m,P)$ and then, conditional on that draw, selecting from $p(\beta | m,P,\sigma)$.
 The posterior distribution for
 $\beta$, conditional on the value of $\sigma$, follows the multivariate normal distribution,
\begin{equation}\label{eqn:betapost}
\beta | m,P,\sigma \sim N(\widehat{\beta}, (\mathbf{X}_*'\Sigma_*^{-1}\mathbf{X}_*)^{-1}),
\end{equation}
where $\widehat{\beta}$ is the standard maximum likelihood (weighted least squares) solution,
$\widehat{\beta} = (\mathbf{X}_*'\Sigma_*^{-1}\mathbf{X}_*)^{-1}\mathbf{X}_*'\Sigma_*^{-1}m_*.$
Unlike the posterior distribution of $\beta$ (given $\sigma$), the posterior distribution of $\sigma$, $p(\sigma^2 | m,P)$, does not follow
a simple conjugate distribution.  Instead, the distribution  follows the form
\begin{equation}
\label{eqn:sigpost}
p(\sigma^2 | m,P) \propto \frac{p(\beta)p(\sigma^2) L(m | P,\beta,\sigma)}{p(\beta | m,P,\sigma)},
\end{equation}
where the prior on $\beta$ is proportional to the informative prior on $\mu$, the flat prior on $\sigma$ is $p(\sigma^2) \propto \sigma^{-2}$, and the data likelihood $L$
is the product, over all  observed magnitudes, of the Gaussian likelihood of the data given the model (Equation \ref{eqn:model}) with all parameters specified.

We  draw samples from our joint posterior distribution  $p(\beta,\sigma | m,P)$ using Equations \ref{eqn:betapost} and 
\ref{eqn:sigpost} in conjunction.  In practice, we compute\footnote{Assuming  that $\beta = \widehat{\beta}$.  Several iterations show that the posterior distribution of $\sigma$ is insensitive to the assumed choice of $\beta$.} $p(\sigma^2|m,P)$ over a fine grid of $\sigma$ values using Equation \ref{eqn:sigpost}, and then draw a sample of $\sigma$  from this density. For each sampled $\sigma$, we subsequently draw a $\beta$ from Equation \ref{eqn:betapost}, conditional on the drawn $\sigma$ value. We repeat this process 10,000 times to characterize the joint posterior distribution. Using a large sample from this joint posterior distribution, we can compute  quantities of interest such as the maximum {\it a posteriori} slopes and zero points of the PL relationship of each WISE band, the intrinsic scatter of the data around the PL relationship in each band, and the spread in the {\it a posteriori} distribution of the PL parameters.

\section{Comparison with Traditional Fit}\label{s:comp}

\begin{figure*} [tb]
 \includegraphics[width=6.85in]{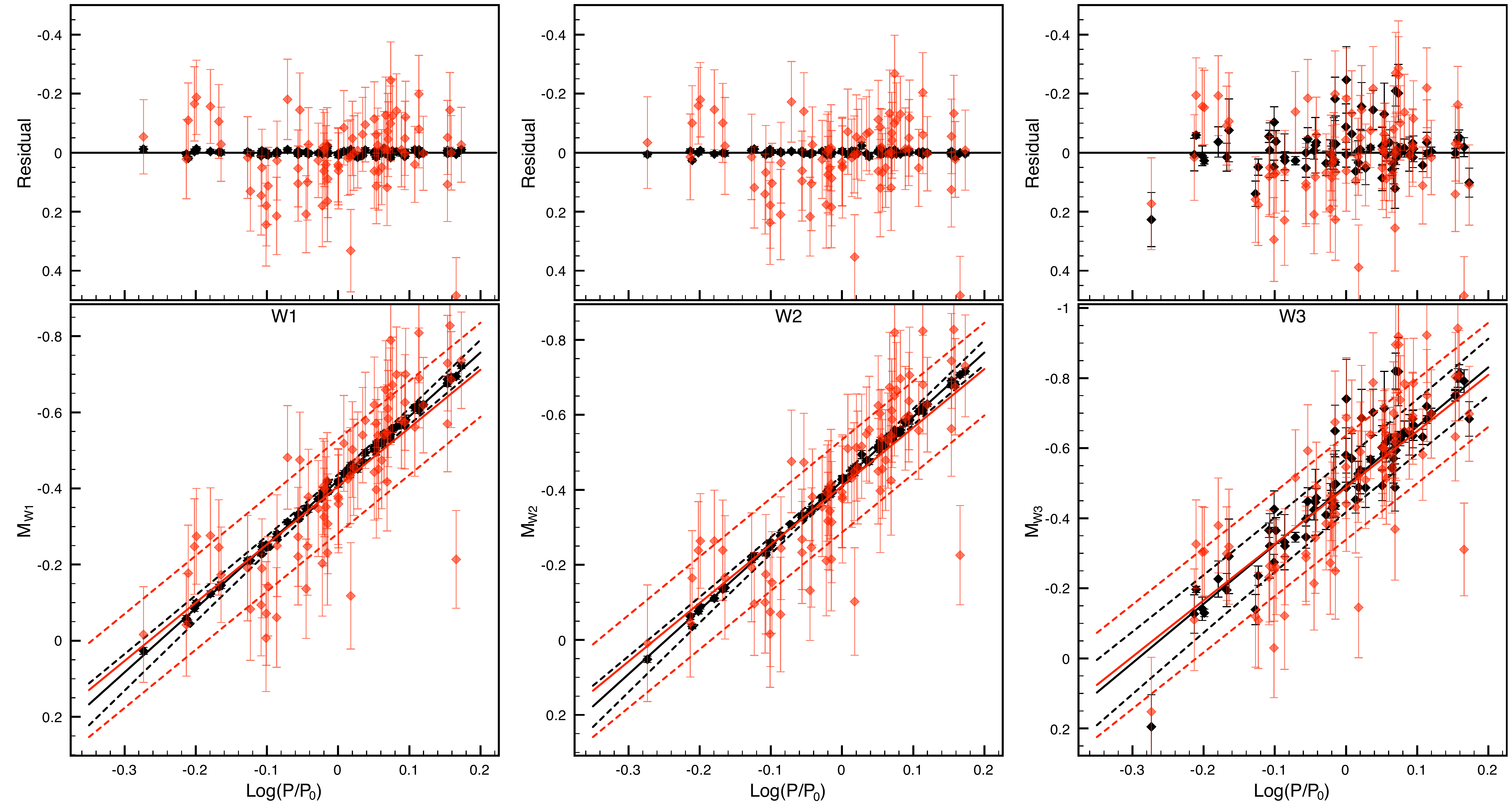}
 \caption{Period-luminosity relations for W1, W2, and W3 (left to right). Results of the Bayesian fitting procedure are shown in black, and the results of the least squares fit are overplotted in red. In each figure, the solid lines show the models' predictions of the RRL absolute magnitude, as a function of RRL period.  The dashed lines show the $\pm 1\sigma$ scatter.  The top panel of each plot shows the residual spread around the best fit model of each fit procedure. The error bars of the Bayesian fit for W1 and W2 are smaller than the diamond markers.} 
 \label{fig:comparison}
 \end{figure*}

To demonstrate the improvement in the fit from the Bayesian approach, which in turn means an improvement in the predicted distances resulting from the calibrated PL relation, we compare it to a traditional least squares regression fit. Using the same prior distances (technically, the expectation value of the prior distance distributions), period measurements, and observed WISE mean flux magnitudes we perform a least squares fit for the slope $\alpha$ and zero point $\beta$ of the PL relation. We calculate the scatter ($1\sigma$) as the standard deviation of the residuals to the fit. The fit parameters and scatter are presented in Table \ref{tbl:comparison}. In Figure \ref{fig:comparison} we overplot in red the least squares PL relation fits into Figure 5 from \cite{2011ApJ...738..185K}. 

 \begin{table}[htb]
\footnotesize
 \caption{Comparison of least squares (subscript LS) and Bayesian (subscript B) fits to the WISE RRL PL relations. We calculate the scatter ($1\sigma$) as the standard deviation of the residuals to each fit.} 
  \label{tbl:comparison}
 \begin{tabular}{|c||c|c|c||c|c|c|} \hline
band & $\alpha_{\rm LS}$ & $\beta_{\rm LS}$ & $1\sigma_{\rm LS}$ &  $\alpha_{\rm B}$ & $\beta_{\rm B}$ & $1\sigma_{\rm B}$ \\ \hline
W1 & -0.420 & -1.675 & 0.124 &  -0.421 & -1.681 & 0.007 \\ \hline
W2 & -0.425 & -1.713 & 0.124 &  -0.423 & -1.715 & 0.007 \\ \hline
W3 & -0.503 & -1.763 & 0.149 &  -0.493 & -1.688 & 0.074 \\ \hline
 \end{tabular}
\end{table}

As expected, the actual parameters of the PL relations, zero point and slope, are statistically consistent. Comparison of their $1\sigma$ scatter, however, illustrates that the Bayesian approach produces a set of PL relations with nearly eighteen times lower scatter in WISE bands W1 and W2 and two times lower scatter in W3. The Bayesian fit's significant reduction in scatter primarily results from allowing the posterior distances to be fit from within the prior distances' distributions. Thus, the scatter of the Bayesian fit more closely approaches the true intrinsic scatter.

\section{Conclusions}\label{s:conc}

Table \ref{tbl:comparison} and Figure \ref{fig:comparison} clearly demonstrate the Bayesian modeling approach's ability to produce a set of more tightly constrained PL relations. This benefit of the approach primarily arrises from its taking into account the full prior distribution on the distances and allowing the fit to refine these distance distributions. The traditional least squares fitting method instead utilizes only the expectation value of these prior distances and leaves them unchanged during the fitting procedure, hence producing a fit with significantly larger scatter.

It is natural to inquire as to whether the Bayesian model is overly constrained. That is, one must be careful not to allow the posterior distance distributions to diverge significantly from the prior distributions. To evaluate this property of the Bayesian fit it is reasonable to compute and examine the prior, posterior, and prediction probability densities, as shown for four examples in Figure 7 of \cite{2011ApJ...738..185K}. Finally, we note that the the newest HST parallax measurements for the four RRL variables V*RRLyr, V*RZCep, V*SUDra, and V*UVOct \citep{2011arXiv1109.5631B} were not yet available for use as distance priors in the Bayesian fit performed in \cite{2011ApJ...738..185K}, however the distance posteriors produced in that work compare quite well with these newly published parallax distances (see Table \ref{tbl:distances}). We interpret these predictions of the parallax distances as a strong vindication of our methodology.

 \begin{table}[htb]
\footnotesize
 \caption{Recently published RRL HST parallax distances \citep{2011arXiv1109.5631B} compared with the values which \cite{2011ApJ...738..185K} previously predicted through Bayesian linear model fitting of the mid-infrared PL relations. All distances are given in parsecs.} 
  \label{tbl:distances}
 \begin{tabular}{|l|c|c|c|} \hline
Name & HST $d$ & PL Fit $d$ & Bayesian Prior $d$ \\ \hline
V*RRLyr & $265\pm 9$ & $253\pm2$ & $262\pm15$ \\ \hline
V*RZCep & $394\pm 30$ & $381\pm6$ & $405\pm23$ \\ \hline
V*SUDra & $704\pm 80$ & $696\pm7$ & $696\pm40$ \\ \hline
V*UVOct & $585\pm 34$ & $536\pm4$ & $553\pm32$ \\ \hline
 \end{tabular}
\end{table}

While it is true that the Bayesian modeling method described in this article is not strictly applicable to all situations (for example, when fitting data which share a common distance prior as is standard for studies of pulsating variables in the Large Magellanic Cloud), it is nevertheless an invaluable tool for calibrating the Galactic PL relations of RRL, Cepheid, and Mira variables. This approach also has the potential to improve the model fits of other phenomena, such as the calibration of other distance indicators (SNe Ia, planetary nebulae, tip of the red giant branch, etc). Our intent with this article is to spread awareness of the benefits and applicability of this Bayesian approach and encourage future PL relation investigations to consider employing this powerful analysis method.

%
\acknowledgments
The authors acknowledge the generous support of a CDI grant (\#0941742) from the National Science Foundation. JSB and CRK were also partially supported by grant NSF/AST-100991.


 \bibliographystyle{spr-mp-nameyear-cnd}  
 \bibliography{Klein_refs}                

\begin{thebibliography}{16}
\ifx \bisbn   \undefined \def \bisbn  #1{ISBN #1}\fi
\ifx \binits  \undefined \def \binits#1{#1} \fi
\ifx \bauthor  \undefined \def \bauthor#1{#1} \fi
\ifx \batitle  \undefined \def \batitle#1{#1} \fi
\ifx \bjtitle  \undefined \def \bjtitle#1{#1}\fi
\ifx \bvolume  \undefined \def \bvolume#1{\textbf{#1}}\fi
\ifx \byear  \undefined \def \byear#1{#1} \fi
\ifx \bissue  \undefined \def \bissue#1{#1} \fi
\ifx \bfpage  \undefined \def \bfpage#1{#1} \fi
\ifx \blpage  \undefined \def \blpage #1{#1} \fi
\ifx \burl  \undefined \def \burl#1{\textsf{#1}} \fi
\ifx \doiurl  \undefined \def \doiurl#1{\textsf{#1}} \fi
\ifx \betal  \undefined \def \betal{\textit{et al.}} \fi
\ifx \binstitute  \undefined \def \binstitute#1{#1} \fi
\ifx \binstitutionaled  \undefined \def \binstitutionaled#1{#1} \fi
\ifx \bctitle  \undefined \def \bctitle#1{#1} \fi
\ifx \beditor  \undefined \def \beditor#1{#1} \fi
\ifx \bpublisher  \undefined \def \bpublisher#1{#1} \fi
\ifx \bbtitle  \undefined \def \bbtitle#1{#1} \fi
\ifx \bedition  \undefined \def \bedition#1{#1} \fi
\ifx \bseriesno  \undefined \def \bseriesno#1{#1} \fi
\ifx \blocation  \undefined \def \blocation#1{#1} \fi
\ifx \bsertitle  \undefined \def \bsertitle#1{#1} \fi
\ifx \bsnm \undefined \def \bsnm#1{#1} \fi
\ifx \bsuffix \undefined \def \bsuffix#1{#1} \fi
\ifx \bparticle \undefined \def \bparticle#1{#1} \fi
\ifx \barticle \undefined \def \barticle#1{#1} \fi
\ifx \bconfdate \undefined \def \bconfdate #1{#1} \fi
\ifx \botherref \undefined \def \botherref #1{#1} \fi
\ifx \url \undefined \def \url#1{\textsf{#1}} \fi
\ifx \bchapter \undefined \def \bchapter#1{#1} \fi
\ifx \bbook \undefined \def \bbook#1{#1} \fi
\ifx \bcomment \undefined \def \bcomment#1{#1} \fi
\ifx \oauthor \undefined \def \oauthor#1{#1} \fi
\ifx \citeauthoryear \undefined \def \citeauthoryear#1{#1} \fi
\ifx \endbibitem  \undefined \def \endbibitem {}\fi
\ifx \bconflocation  \undefined \def \bconflocation#1{#1} \fi
\ifx \arxivurl  \undefined \def \arxivurl#1{\textsf{#1}} \fi

\bibitem[\protect\citeauthoryear{{Barnes} et~al.}{2003}]{2003ApJ...592..539B}
\begin{barticle}
\bauthor{\bsnm{{Barnes}}, \binits{T.G.} \bsuffix{III}},
\bauthor{\bsnm{{Jefferys}}, \binits{W.H.}},
\bauthor{\bsnm{{Berger}}, \binits{J.O.}},
\bauthor{\bsnm{{Mueller}}, \binits{P.J.}},
\bauthor{\bsnm{{Orr}}, \binits{K.}},
\bauthor{\bsnm{{Rodriguez}}, \binits{R.}}:
\bjtitle{\apj}
\bvolume{592},
\bfpage{539}
(\byear{2003}).
\arxivurl{arXiv:astro-ph/0303656}.
doi:\doiurl{10.1086/375583}
\end{barticle}
\endbibitem

\bibitem[\protect\citeauthoryear{{Benedict} et~al.}{2007}]{2007AJ....133.1810B}
\begin{barticle}
\bauthor{\bsnm{{Benedict}}, \binits{G.F.}},
\bauthor{\bsnm{{McArthur}}, \binits{B.E.}},
\bauthor{\bsnm{{Feast}}, \binits{M.W.}},
\bauthor{\bsnm{{Barnes}}, \binits{T.G.}},
\bauthor{\bsnm{{Harrison}}, \binits{T.E.}},
\bauthor{\bsnm{{Patterson}}, \binits{R.J.}},
\bauthor{\bsnm{{Menzies}}, \binits{J.W.}},
\bauthor{\bsnm{{Bean}}, \binits{J.L.}},
\bauthor{\bsnm{{Freedman}}, \binits{W.L.}}:
\bjtitle{\aj}
\bvolume{133},
\bfpage{1810}
(\byear{2007}).
\arxivurl{arXiv:astro-ph/0612465}.
doi:\doiurl{10.1086/511980}
\end{barticle}
\endbibitem

\bibitem[\protect\citeauthoryear{{Benedict} et~al.}{2011}]{2011AJ....142..187B}
\begin{barticle}
\bauthor{\bsnm{{Benedict}}, \binits{G.F.}},
\bauthor{\bsnm{{McArthur}}, \binits{B.E.}},
\bauthor{\bsnm{{Feast}}, \binits{M.W.}},
\bauthor{\bsnm{{Barnes}}, \binits{T.G.}},
\bauthor{\bsnm{{Harrison}}, \binits{T.E.}},
\bauthor{\bsnm{{Bean}}, \binits{J.L.}},
\bauthor{\bsnm{{Menzies}}, \binits{J.W.}},
\bauthor{\bsnm{{Chaboyer}}, \binits{B.}},
\bauthor{\bsnm{{Fossati}}, \binits{L.}},
\bauthor{\bsnm{{Nesvacil}}, \binits{N.}},
\bauthor{\bsnm{{Smith}}, \binits{H.A.}},
\bauthor{\bsnm{{Kolenberg}}, \binits{K.}},
\bauthor{\bsnm{{Laney}}, \binits{C.D.}},
\bauthor{\bsnm{{Kochukhov}}, \binits{O.}},
\bauthor{\bsnm{{Nelan}}, \binits{E.P.}},
\bauthor{\bsnm{{Shulyak}}, \binits{D.V.}},
\bauthor{\bsnm{{Taylor}}, \binits{D.}},
\bauthor{\bsnm{{Freedman}}, \binits{W.L.}}:
\bjtitle{\aj}
\bvolume{142},
\bfpage{187}
(\byear{2011}).
\arxivurl{1109.5631}.
doi:\doiurl{10.1088/0004-6256/142/6/187}
\end{barticle}
\endbibitem

\bibitem[\protect\citeauthoryear{{Cacciari}}{2009}]{2009MmSAI..80...97C}
\begin{barticle}
\bauthor{\bsnm{{Cacciari}}, \binits{C.}}:
\bjtitle{\memsai}
\bvolume{80},
\bfpage{97}
(\byear{2009})
\end{barticle}
\endbibitem

\bibitem[\protect\citeauthoryear{{Chaboyer}}{1999}]{postHipp..book}
\begin{bbook}
\bauthor{\bsnm{{Chaboyer}}, \binits{B.}}:
\bbtitle{{Post-Hipparcos Cosmic Candles}}.
\bsertitle{1},
vol. \bseriesno{111}.
\bpublisher{{Dordrecht: Kluwer}}, \blocation{???}
(\byear{1999})
\end{bbook}
\endbibitem

\bibitem[\protect\citeauthoryear{{Feast} and
  {Catchpole}}{1997}]{1997MNRAS.286L...1F}
\begin{barticle}
\bauthor{\bsnm{{Feast}}, \binits{M.W.}},
\bauthor{\bsnm{{Catchpole}}, \binits{R.M.}}:
\bjtitle{\mnras}
\bvolume{286},
\bfpage{1}
(\byear{1997})
\end{barticle}
\endbibitem

\bibitem[\protect\citeauthoryear{{Gelman} et~al.}{{2003}}]{gelman2003}
\begin{bbook}
\bauthor{\bsnm{{Gelman}}, \binits{A.}},
\bauthor{\bsnm{{Carlin}}, \binits{J.B.}},
\bauthor{\bsnm{{Stern}}, \binits{H.S.}},
\bauthor{\bsnm{{Rubin}}, \binits{D.B.}}:
\bbtitle{{Bayesian Data Analysis, Second Edition}},
\bedition{{2}} edn.
\bpublisher{{Chapman and Hall/CRC}}, \blocation{???}
(\byear{{2003}})
\end{bbook}
\endbibitem

\bibitem[\protect\citeauthoryear{{Glass} and
  {Evans}}{2003}]{2003MNRAS.343...67G}
\begin{barticle}
\bauthor{\bsnm{{Glass}}, \binits{I.S.}},
\bauthor{\bsnm{{Evans}}, \binits{T.L.}}:
\bjtitle{\mnras}
\bvolume{343},
\bfpage{67}
(\byear{2003}).
doi:\doiurl{10.1046/j.1365-8711.2003.06632.x}
\end{barticle}
\endbibitem

\bibitem[\protect\citeauthoryear{{Klein} et~al.}{2011}]{2011ApJ...738..185K}
\begin{barticle}
\bauthor{\bsnm{{Klein}}, \binits{C.R.}},
\bauthor{\bsnm{{Richards}}, \binits{J.W.}},
\bauthor{\bsnm{{Butler}}, \binits{N.R.}},
\bauthor{\bsnm{{Bloom}}, \binits{J.S.}}:
\bjtitle{\apj}
\bvolume{738},
\bfpage{185}
(\byear{2011}).
\arxivurl{1105.0055}.
doi:\doiurl{10.1088/0004-637X/738/2/185}
\end{barticle}
\endbibitem

\bibitem[\protect\citeauthoryear{{M.A.C.~Perryman \&
  ESA}}{1997}]{1997ESASP1200.....P}
\begin{bbook}
\beditor{\bsnm{{M.A.C.~Perryman \& ESA}}} (ed.):
\bbtitle{{The Hipparcos and Tycho Catalogues. Astrometric and Photometric Star
  Catalogues Derived from the Esa Hipparcos Space Astrometry Mission}}.
\bsertitle{ESA Special Publication},
vol. \bseriesno{1200}
(\byear{1997})
\end{bbook}
\endbibitem

\bibitem[\protect\citeauthoryear{{Matsunaga}
  et~al.}{2006}]{2006MNRAS.370.1979M}
\begin{barticle}
\bauthor{\bsnm{{Matsunaga}}, \binits{N.}},
\bauthor{\bsnm{{Fukushi}}, \binits{H.}},
\bauthor{\bsnm{{Nakada}}, \binits{Y.}},
\bauthor{\bsnm{{Tanab{\'e}}}, \binits{T.}},
\bauthor{\bsnm{{Feast}}, \binits{M.W.}},
\bauthor{\bsnm{{Menzies}}, \binits{J.W.}},
\bauthor{\bsnm{{Ita}}, \binits{Y.}},
\bauthor{\bsnm{{Nishiyama}}, \binits{S.}},
\bauthor{\bsnm{{Baba}}, \binits{D.}},
\bauthor{\bsnm{{Naoi}}, \binits{T.}},
\bauthor{\bsnm{{Nakaya}}, \binits{H.}},
\bauthor{\bsnm{{Kawadu}}, \binits{T.}},
\bauthor{\bsnm{{Ishihara}}, \binits{A.}},
\bauthor{\bsnm{{Kato}}, \binits{D.}}:
\bjtitle{\mnras}
\bvolume{370},
\bfpage{1979}
(\byear{2006}).
\arxivurl{arXiv:astro-ph/0606609}.
doi:\doiurl{10.1111/j.1365-2966.2006.10620.x}
\end{barticle}
\endbibitem

\bibitem[\protect\citeauthoryear{{Riess} et~al.}{2011}]{2011ApJ...730..119R}
\begin{barticle}
\bauthor{\bsnm{{Riess}}, \binits{A.G.}},
\bauthor{\bsnm{{Macri}}, \binits{L.}},
\bauthor{\bsnm{{Casertano}}, \binits{S.}},
\bauthor{\bsnm{{Lampeitl}}, \binits{H.}},
\bauthor{\bsnm{{Ferguson}}, \binits{H.C.}},
\bauthor{\bsnm{{Filippenko}}, \binits{A.V.}},
\bauthor{\bsnm{{Jha}}, \binits{S.W.}},
\bauthor{\bsnm{{Li}}, \binits{W.}},
\bauthor{\bsnm{{Chornock}}, \binits{R.}}:
\bjtitle{\apj}
\bvolume{730},
\bfpage{119}
(\byear{2011}).
\arxivurl{1103.2976}.
doi:\doiurl{10.1088/0004-637X/730/2/119}
\end{barticle}
\endbibitem

\bibitem[\protect\citeauthoryear{{Sollima} et~al.}{2006}]{2006MNRAS.372.1675S}
\begin{barticle}
\bauthor{\bsnm{{Sollima}}, \binits{A.}},
\bauthor{\bsnm{{Cacciari}}, \binits{C.}},
\bauthor{\bsnm{{Valenti}}, \binits{E.}}:
\bjtitle{\mnras}
\bvolume{372},
\bfpage{1675}
(\byear{2006}).
\arxivurl{arXiv:astro-ph/0608397}.
doi:\doiurl{10.1111/j.1365-2966.2006.10962.x}
\end{barticle}
\endbibitem

\bibitem[\protect\citeauthoryear{{van Leeuwen}}{2007}]{2007ASSL..350.....V}
\begin{bbook}
\beditor{\bsnm{{van Leeuwen}}, \binits{F.}} (ed.):
\bbtitle{{Hipparcos, the New Reduction of the Raw Data}}.
\bsertitle{Astrophysics and Space Science Library},
vol. \bseriesno{350}
(\byear{2007})
\end{bbook}
\endbibitem

\bibitem[\protect\citeauthoryear{{Whitelock}
  et~al.}{2008}]{2008MNRAS.386..313W}
\begin{barticle}
\bauthor{\bsnm{{Whitelock}}, \binits{P.A.}},
\bauthor{\bsnm{{Feast}}, \binits{M.W.}},
\bauthor{\bsnm{{van Leeuwen}}, \binits{F.}}:
\bjtitle{\mnras}
\bvolume{386},
\bfpage{313}
(\byear{2008}).
\arxivurl{0801.4465}.
doi:\doiurl{10.1111/j.1365-2966.2008.13032.x}
\end{barticle}
\endbibitem

\bibitem[\protect\citeauthoryear{{Wright} et~al.}{2010}]{2010AJ....140.1868W}
\begin{barticle}
\bauthor{\bsnm{{Wright}}, \binits{E.L.}},
\bauthor{\bsnm{{Eisenhardt}}, \binits{P.R.M.}},
\bauthor{\bsnm{{Mainzer}}, \binits{A.K.}},
\bauthor{\bsnm{{Ressler}}, \binits{M.E.}},
\bauthor{\bsnm{{Cutri}}, \binits{R.M.}},
\bauthor{\bsnm{{Jarrett}}, \binits{T.}},
\bauthor{\bsnm{{Kirkpatrick}}, \binits{J.D.}},
\bauthor{\bsnm{{Padgett}}, \binits{D.}},
\bauthor{\bsnm{{McMillan}}, \binits{R.S.}},
\bauthor{\bsnm{{Skrutskie}}, \binits{M.}},
\bauthor{\bsnm{{Stanford}}, \binits{S.A.}},
\bauthor{\bsnm{{Cohen}}, \binits{M.}},
\bauthor{\bsnm{{Walker}}, \binits{R.G.}},
\bauthor{\bsnm{{Mather}}, \binits{J.C.}},
\bauthor{\bsnm{{Leisawitz}}, \binits{D.}},
\bauthor{\bsnm{{Gautier}}, \binits{T.N.}},
\bauthor{\bsnm{{McLean}}, \binits{I.}},
\bauthor{\bsnm{{Benford}}, \binits{D.}},
\bauthor{\bsnm{{Lonsdale}}, \binits{C.J.}},
\bauthor{\bsnm{{Blain}}, \binits{A.}},
\bauthor{\bsnm{{Mendez}}, \binits{B.}},
\bauthor{\bsnm{{Irace}}, \binits{W.R.}},
\bauthor{\bsnm{{Duval}}, \binits{V.}},
\bauthor{\bsnm{{Liu}}, \binits{F.}},
\bauthor{\bsnm{{Royer}}, \binits{D.}},
\bauthor{\bsnm{{Heinrichsen}}, \binits{I.}},
\bauthor{\bsnm{{Howard}}, \binits{J.}},
\bauthor{\bsnm{{Shannon}}, \binits{M.}},
\bauthor{\bsnm{{Kendall}}, \binits{M.}},
\bauthor{\bsnm{{Walsh}}, \binits{A.L.}},
\bauthor{\bsnm{{Larsen}}, \binits{M.}},
\bauthor{\bsnm{{Cardon}}, \binits{J.G.}},
\bauthor{\bsnm{{Schick}}, \binits{S.}},
\bauthor{\bsnm{{Schwalm}}, \binits{M.}},
\bauthor{\bsnm{{Abid}}, \binits{M.}},
\bauthor{\bsnm{{Fabinsky}}, \binits{B.}},
\bauthor{\bsnm{{Naes}}, \binits{L.}},
\bauthor{\bsnm{{Tsai}}, \binits{C.}}:
\bjtitle{\aj}
\bvolume{140},
\bfpage{1868}
(\byear{2010}).
\arxivurl{1008.0031}.
doi:\doiurl{10.1088/0004-6256/140/6/1868}
\end{barticle}
\endbibitem

\end{thebibliography}

\end{document}